\documentclass[12pt,a4paper]{article}
\pdfoutput=1
\usepackage{jheppub}
\usepackage[T1]{fontenc} 
\usepackage{epsfig} 
\usepackage{amscd}
\usepackage{verbatim}
\usepackage{latexsym}
\usepackage{amsfonts,amsthm,amsmath,mathtools}
\usepackage{color}

\notoc 

\title{\centering{3D $\tau_{RR}$-minimization\\  in \\AdS$_4$ gauged  supergravity}}
\preprint{CCNY-HEP-15/6}
\author[a]{Antonio Amariti}
\author[b]{\!\!,~Alessandra Gnecchi}
\affiliation[a]{Physics Department, The City College of the CUNY,
160 Convent Avenue, New York, NY 10031, USA}
\affiliation[b]{Institute for Theoretical Physics, KU Leuven, 3001 Leuven, Belgium}
\usepackage{amssymb}
\usepackage{appendix}
\usepackage[utf8]{inputenc}
\usepackage[english]{babel}
\usepackage{amsmath, amssymb, graphics}
\usepackage{amsfonts, color}
\usepackage{mathrsfs}

\emailAdd{aamariti@ccny.cuny.edu,alessandra.gnecchi@fys.kuleuven.be}

\abstract{In this paper we propose the identification in
AdS$_4$ $\mathcal{N}=2$ gauged supergravity
of
the coefficient  $\tau_{RR}$ of  3D $\mathcal{N}=2$ SCFTs.  
We constraint the  structure of this function in supergravity
by combining the results from unitarity, holography and localization.
We show that our conjectured function is minimized by the exact R-charge,
corresponding to a gravitational attractor for the scalars in the special geometry.
We identify this mechanism with the supergravity dual of the $\tau_{RR}$-minimization.
We check this proposal in the  ABJM model, comparing with the expectations from localization and the AdS/CFT duality.
We comment also on some possible relations with the black hole 
microstate counting, recently obtained from the application of localization techniques.}

\begin{document}

\maketitle

\newpage
\tableofcontents

\section{Overview}

The comprehension of the mathematical structure of the renormalization group (RG)
flow is one of the main goals of modern theoretical physics.
A general expectation is that the RG flow is irreversible and the
degrees of freedom reduce when an ultraviolet (UV)
theory flows to the infrared (IR).
There have been many proposals for quantifying this idea.
In even dimensions 
exact results have been obtained with the aid of
global anomalies.
The existence of a monotonically decreasing function interpolating between the UV and the IR fixed points  of an RG flow has been proven in 2D \cite{Zamolodchikov:1986gt}  and 4D 
\cite{Komargodski:2011vj}.
In 2D  this function coincides, at the 
UV and at the IR fixed points, with the Weyl anomaly, the central charge $c$. In 4D the role is played by the coefficient of the Euler density, the central charge $a$. These identifications led to the names of $c$-theorem in 2D and $a$-theorem in 4D.

In supersymmetric field theories the central charge $a$
has been  computed non-perturbatively from the current correlators of the three point 
functions \cite{Anselmi:1997am}, and it is
\begin{equation}
a = \frac{3}{32} \left( \text{Tr} R^3 - \text{Tr} R \right)
\end{equation}
where R represents the charge associated to  the $U(1)_R$ symmetry.
In the superconformal case R is related to the scaling dimension $\Delta$.
This relation is a consequence of the fact that the R-current
is the lowest component of the stress tensor supermultiplet.
This current is not in general the UV R-current $J_0$, because 
during an RG flow $J_0$ mixes with the other global abelian flavor symmetries  $F_i$.
At the fixed point the exact R-current becomes
\begin{equation}
J_R = J_0 + \delta_i F^i
\end{equation}
The  current $J_R $ corresponds to the choice of the $\delta_i$ coefficients
\footnote{Observe that the coefficient of the current $J_0$ depends of the normalization
of the $R$-current. Here we discuss the canonical normalization. In general the difference 
between an R-current and a non-R-current is the fact that the superpotential is charged under the former
but not under the latter.} maximizing the 
central charge $a$  \cite{Intriligator:2003jj}. This principle has been named $a$-maximization.

Superconformal field theories received a large attention because of their role in the holographic correspondence. Indeed the correlators of the strongly coupled $d$-dimensional CFT can be computed at the boundary of the $d+1$ dimensional dual  theory.
The correspondence allowed to identify the central charge with the superpotential
of the dual $\mathcal{N}=2$  AdS$_5$ gauged supergravity theory \cite{Tachikawa:2005tq}.
The key role in this identification was played by the Chern-Simons coefficients of the 
gauged 5d theory.  
The holographic dictionary translated them into the anomalous coefficients of the three point functions of the global currents on the field theory side.
This observation, and the identification of the field theory R-charges with the scalars in the 
very special geometry of  $\mathcal{N}=2$  AdS$_5$ gauged supergravity,
allowed to reformulate  $a$-maximization in this language.

A similar situation has been worked out in the last years in 3D.
Here the situation is more complicate, because of the absence of global anomalies.
More involved techniques are necessary to test the ideas about the irreversibility of the RG flow. Also the counterpart of $a$-maximization is not immediately obvious.

This last problem was first solved in \cite{Barnes:2005bm}, where it has been shown that 
the coefficient $C_T$ of the  two point functions
for the stress energy can determine the exact R-current.
This quantity is related by supersymmetry to the coefficient $\tau_{RR}$
of the two point function of the R-current. 
It has been shown that $\tau_{RR}$ is minimized by the exact R-current.
This result holds also in 4D as a corollary of $a$-maximization. 
Despite the simplicity of this relation the analysis of this quantity is difficult because
it cannot be extracted from non perturbative analysis, making the method rather inefficient.

The breakthrough has  then been achieved by the application of  localization techniques. 
It has been shown indeed that the 
free energy computed on $S^3$, $F_{S^3}$,  is  extremized by the exact  R-current
\cite{Jafferis:2010un}.
By squashing this manifold it has been proved that the free energy is also maximized by the exact R-charge, essentially
for the same reason for which $\tau_{RR}$ is minimized \cite{Closset:2012vg}.
At the fixed point it has also been shown that $\tau_{RR}$ and the free energy are proportional
\cite{Closset:2012ru}, 
even though the general functional relation in terms of the mixing with the R-charge is not known.

The parallelism with the 4D case naturally turned the interest to the supergravity side.
In this case the calculation of $F_{S^3}$ requires 
a holographic calculation in the  AdS$_4$ space with Euclidean signature. 
This analysis has been performed in \cite{Freedman:2013ryh}
and the large $N$ behavior of the ABJM model has been reproduced.
More recently a non perturbative
analysis of the dual mechanism of localization in holography
has been performed in \cite{Dabholkar:2014wpa}.

The relation between the $\tau_{RR}$-function and
 the AdS$_4$ case with Lorentzian signature
has nevertheless been overlooked.  
It has been shown that one can obtain the $\tau_{RR}$-function in AdS$_4$ gauged 
supergravity by consistent truncation of $M$-theory on SE$_7$  manifolds \cite{Barnes:2005bw},
but the approach requires the knowledge of the full 10D geometry.

In this paper we initiate the study of the $\tau_{RR}$-function in $\mathcal{N}=2$
AdS$_4$ gauged supergravity, in presence of a generic (dyonic) gauging. 
We use the holographic dictionary to associate the relevant field theory quantities,
the charges and the currents, 
to the scalar fields and the gauge coupling in the supergravity description.
This dictionary allows us to propose the supergravity dual $\tau_{RR}$-function.
We show that its minimization corresponds to the attractor mechanism for the scalars
of the special geometry, while the minimization corresponds to requirement of positivity
for the metric  on the special manifold.
By including the hypermultiplets in the analysis we further constraint the set of charges involved in the minimization. 
They provide a counterpart of the constraints imposed by the superpotential interactions in the dual field theory.

The paper is organized as follows.
In section \ref{section:tau} we review the field theory aspects of $\tau_{RR}$ maximization that will be useful 
for our discussion. We comment on the fixed point relation between the $\tau_{RR}$-function and $F_{S^3}$ computed at large $N$.
In section \ref{sectionads4} we review some basics aspects of gauged supergravity that will be relevant in our derivation of the holographic dual of $\tau_{RR}$-minimization.
Section \ref{section:main} contains the main result of the paper. We identify 
the holographic dual $\tau_{RR}$-function and show its minimization from gauged supergravity.
This is done by identifying the R-current from the combination of vector fields that appear in the
gravitino variation. 
This corresponds to a combination of the constrained scalars in the special geometry with the massless gauge fields. 
The exact R-current is the
combination corresponding to the graviphoton of the $\mathcal{N}=2$ vacuum. 
By combining the constraint from the special geometry, the holographic dictionary 
and the results from localization we identify the supergravity dual $\tau_{RR}$.
It corresponds to the quartic power of the superpotential appearing in the fermionic variations. 
In section \ref{section:examples} we discuss the ABJM model to show the whole procedure at work.
In section \ref{volufree} we compare the off-shell behavior of the $\tau_{RR}$-function of the ABJM model
expected from AdS/CFT duality,
and discute the possibility of a general relation at large $N$ between the function $\tau_{RR}$ and the free energy,
in terms of a generic assignment of R-charges. 
In section \ref{zaffa} we emphasize the possible connection with the 
recent counting of microstates of 2D AdS$_2$ black holes
and an 1D R-charge extremization principle .
In section \ref{section:discussion} we conclude.
In appendix \ref{appeale} we give further details of AdS$_4$ gauged supergravity.
\section{$\tau_{RR}$-minimization}
\label{section:tau}
In this section we review some aspects of $\tau_{RR}$-minimization
as discussed in \cite{Barnes:2005bm}.
The $\tau_{RR}$-minimization is a method to obtain the exact 
R-charge of a superconformal field theory among all the possible choices allowed by the superconformal  algebra.
The proof of this statement is based on the analysis of the correlation functions of
the two point global  currents 
in superconformal field theories:
\begin{equation}
\langle
j_i^{\mu} (x) j_j^{\nu}(y) 
\rangle
= 
\frac{\tau_{ij}}{2 \pi^3} 
(\partial^2 \delta^{\mu \nu} - \partial^{\mu \nu})
\frac{1}{(x-y)^2}
\end{equation}
where, because of unitarity, the matrix $\tau_{ij}$ has positive eigenvalues.
In superconformal field theories one of these global currents 
corresponds to the lowest component of the supermultiplet
having the stress energy tensor as highest component.
This is commonly referred as the R-current.

This current is in general a combination of the UV R-current $R_0$
and 
other flavor current $F_i$.
Defining a trial R-current 
\begin{equation}
\label{combo}
R_t = R_0 + \delta_i F_i
\end{equation}
The exact R-current corresponds to a specific assignment of the coefficients $\delta_i$.
Some combinations in (\ref{combo}) are usually excluded by the structure of the interactions
but this is not enough, in general, to fix the coefficients $\delta_i$.
A closer look at the correlation functions imposes the necessary constraints. 
The coefficient  $\tau_{R_t R_t}$ is
\begin{equation}
\tau_{R_t R_t} = 
\tau_{R_0 R_0}
+
2 \sum \delta_i \tau_{R_0 i}
 +
\sum_{i,j} \delta_i \delta_j \tau_{ij}
\end{equation}
This function is minimized by the choice of coefficients $\delta_i$ that correspond to the 
exact R-current.
This has been proved by studying the first and the second derivatives of $\tau_{RR}$
in the $\delta_i$ space.
They correspond to $\tau_{R_0i}$ and $\tau_{ij}$. The first is set to zero by 
supersymmetry and imposes an extremization condition.
The minimization condition is imposed by the unitarity of the matrix $\tau_{ij}$.

Despite the simplicity of this result the $\tau_{RR}$  extremization 
did not become a popular method to extract the R-charge, because of the absence of anomalies in 3D. The general the non-perturbative structure of the $\tau_{RR}$-function is not known
and the function can be used to compute the exact R-charge only at weak coupling in the perturbative expansion.
A more tractable object is the free energy $F_{S^3}$, 
computed by localization of the path integral on $S^3$
\cite{Kapustin:2009kz,Jafferis:2010un,Hama:2010av}. 
This led to a non perturbative exact result and it was shown that 
the $F_{S^3}$ is maximized by the exact R-charge
\cite{Closset:2012vg}. Moreover the non perturbative nature of this 
function allowed the conjecture of a 3D $F$-theorem
\cite{Jafferis:2011zi}.
On the contrary counterexamples to a $\tau_{RR}$ theorem have been
found in \cite{Nishioka:2013gza}.

When considering large $N$ supersymmetric gauge theories 
it has been shown that at the fixed point 
the large $N$ free energy and $\tau_{RR}$  are related by a simple relation
\cite{Closset:2012ru}
\begin{equation}
\label{large}
F_{S^3}^{max} = \frac{\pi^2}{4} \tau_{RR}^{min}
\end{equation} 
In the rest of the paper we look for the $\tau_{RR}$-function 
and its minimization principle from the point of view of
$\mathcal{N}=2$ AdS$_4$ gauged  supergravity.

\section{AdS$_4$ gauged supergravity}
\label{sectionads4}
In this section we review some general aspects of AdS$_4$ gauged supergravity.
This allows us to fix the notations necessary for the rest of the analysis.
We provide  further details in Appendix \ref{appeale}.

At the anti de Sitter vacuum fermionic fields are set to zero, thus the relevant dynamics is given by the bosonic part of the action. In $\mathcal{N}=2$ Supergravity, coupled to $n_V$ vector multiplets and $n_H$ hypermultiplets, it can be written as 
\begin{eqnarray}\label{SGaction}
S&=&\int d^4 x\left(- \frac R2+\mathcal{I}_{\Lambda \Sigma}F^{\Lambda}_{\mu \nu}F^{\Sigma\,\mu\nu} +\frac1{2\sqrt{-g}}\mathcal{R}_{\Lambda \Sigma}\epsilon^{\mu \nu \rho \sigma}F^{\Lambda}_{\mu \nu}F^{\Sigma}_{\rho \sigma}+\right.\nonumber\\
&&+ \left.g_{i\bar{j}}\partial_{\mu}z^i\partial^{\mu}\bar z^{\bar j}+h_{uv}\nabla_\mu q^u\nabla^\mu q^v-V_g(z,\bar z,q)\right)
\end{eqnarray}
Our study will deal with abelian gaugings, in particular where only the scalars of the quaternionic manifold are charged under the gauge fields, while the scalars of the vector multiplet remain neutral. The latter corresponds to the request that only isometries of the Quaternionic manifolds are gauged, namely that the potential is of the form
\begin{eqnarray}
V_g(z,\bar z, q)&=&4 h_{uv}\langle k^u(q),\,\mathcal{V}(z,\bar z) \rangle\, \langle k^u(q),\,\overline{\mathcal{V}}(z,\bar z)\rangle
-3 W\overline{W}+g^{i\bar j}D_i W D_{\bar j}\overline{W}\  
\end{eqnarray}
We are introducing here a complex function built from the product of the moment maps with the symplectic sections as\footnote{We will generically consider both electric and magnetic gauging, thus we indicate
\begin{eqnarray}
\mathcal{P}^x=\mathcal{P}^x_\lambda(\Theta^{\Lambda \lambda},\Theta^\lambda_\Lambda)\ ,
\end{eqnarray}
for a generic choice of embedding tensor $\Theta$ \cite{deWit:2011gk}. Analogously the Killing vectors will be given in a symplectic vector $k^u=(k^{u \Lambda}(q),k^u_\Lambda(q))$.}
\begin{eqnarray}\label{W-def}
W(z,\bar z,q)&=& \langle \mathcal{P}^x(q),\, \mathcal{V}(z,\bar z)\rangle\equiv e^{K/2}\left(\mathcal P^x_{\Lambda}X^{\Lambda}-\mathcal P^{x\Lambda} F_{\Lambda}\right) \ 
\end{eqnarray}
that reduces to the domain walls superpotential in the case of $U(1)$ R-symmetry gauging (Fayet-Iliopoulos).
For a complete account of definitions of special geometry and gauged supergravity we refer to the appendix.

A configuration with zero fermions and zero gauge fields is supersymmetric if the corresponding supersymmetry variations of the fermions vanish. They involve the scalar fields and are explicitely given by
\begin{eqnarray}
\label{Susyvar}
\delta \psi_\mu^A&=&D_\mu \epsilon^A-\frac12(\sigma^x)^A_B \gamma_\mu W \epsilon^B+...
\nonumber\\
\delta \lambda^{iA}&=&i g^{i\bar j}(\sigma^x)^A_B  D_{\bar j}\overline W\epsilon^B+...
\nonumber\\
\delta \zeta^\alpha&=&\mathcal{U}^{A}_{u \alpha}\langle k^u,\overline{\mathcal{V}}\rangle \epsilon_A+..
\end{eqnarray}
where the dots indicate terms which are identically zero at the vacuum.
In all the cases we consider we are always able to use $SU(2)$ symmetry to rotate the moment maps $\mathcal{P}^x$ in the direction $\mathcal{P}^3$, with $\mathcal{P}^1=\mathcal{P}^2=0$ (in particular this is how definition \eqref{W-def} has to be read). 
Notice that there are cases where this is not possible \cite{Guarino:2015jca,Guarino:2015qaa}. This requires modifications in the analysis and we will leave this point to further investigations.

The conditions for the supersymmetric vacuum are then
\begin{eqnarray}
\label{vacuum}
\partial_i|W|\,\big|_{\{q^{u*},\, z^{i*},\, \bar z^{i*}\}}&=&0\ , 
\qquad 
\langle k^u(q^*),e^{-K/2}\mathcal{V}(z^*)\rangle=0\ 
\end{eqnarray}
where the extremization is done only over the scalars $z^i$ of the Special K\"ahler manifold, and $\{q^{u*},\, z^{i*},\, \bar z^{i*}\}$ indicate the value of the scalar fields at the minimum. The other condition depends on how many Killing vectors are identically zero at the vacuum. In particular, the non zero Killing vectors corresponds to a number $n_c$ of algebraic, holomorphic constraints on the fields $z^i$, related to the Higgsing of $n_c$ abelian vector fields at the vacuum \cite{Louis:2012ux}.
Moreover, because of special geometry, if the supersymmetric vacuum is given by the scalar configuration $(q^*,z^*,\bar z^*)$, the $AdS_4$ radius, and thus the cosmological constant at the minimum of the potential can be expressed as
\begin{eqnarray}
\label{moments}
-\frac{\Lambda}{3} &=&\frac{1}{\ell^2_{AdS}}= -\frac12 \mathcal P^{x\,T}(q^*)\mathcal{M}(z^*,\bar z^*)\mathcal{P}^x(q^*)
\end{eqnarray}
by the scalar dependent $Sp(2n_V+2)$ matrix 
\begin{eqnarray}
\label{Male}
\mathcal{M}(z^i,\bar z^{\bar i})=\left(
\begin{array}{cc}
\mathcal{I}+\mathcal{R}\mathcal{I}^{-1}\mathcal{R}&\ \  -\mathcal{R} \mathcal{I}^{-1}\\
-\mathcal{I}^{-1}\mathcal{R}&\mathcal{I}^{-1}
\end{array}
\right)\ 
\end{eqnarray}
where $\mathcal{I}\equiv\mathcal{I}_{\Lambda \Sigma}$ and $\mathcal{R}\equiv\mathcal{R}_{\Lambda \Sigma}$ are the matrices of the scalar  coupling to the gauge fields as taken from the Lagrangian in \eqref{SGaction}. It is important to notice that at the extremum \eqref{vacuum}, the inverse of AdS length square is given by the square root of the quartic symplectic invariant $\mathcal I_4(\mathcal G)$ \cite{Ferrara:1997uz,Andrianopoli:2006ub} valued at the \lq\lq{}charges\rq\rq{} given by the moment maps evaluated at the vacuum as
\begin{eqnarray}
\mathcal G\equiv \mathcal P^x(q^*) \ .
\end{eqnarray}
This follows immediately from the study of black hole horizons attractors \cite{Ferrara:1995ih,Ferrara:1996um,Andrianopoli:1997pn}, where exactly the same extremization occurs with respect to the scalars of the vector multiplets $z^i$. In that case the charges $\mathcal{G}$ are the black hole charges and the quartic invariant gives the value of the black hole entropy, or better, the volume of the $S^2$ at the horizon (black hole area).

\section{$\tau_{RR}$ in $\mathcal{N}=2$ gauged supergravity}
\label{section:main}
In this section we study the $\tau_{RR}$-minimization
from gauged supergravity. 
We start our analysis with the identification of the conserved currents and of the charges that determine their mixing. 
Observe that there can be other 
broken global currents, that we ignore in this first part of the discussion. 
In other words we first restrict the analysis to the case in which the second equation in
(\ref{vacuum}) is solved by setting $k^u$ to zero.
This corresponds of restricting our attention to the sector of the conserved currents
that mix with the $R$-charge, with constant moment maps.

The photons appearing in the supersymmetric variations of the gravitino and of the gaugino are identified with the 
$R$ and the conserved global currents of the dual field theory.
In general the photons that appear in the supersymmetric variations of the fermions are 
combined with the superpotential. In the case of the $R$-current the combination is 
proportional to the superpotential while in the case of the gaugino there is also a derivative involved. 
This distinguishes the 
$R$ current from the other conserved global currents.
In the dual field theory this translates into the fact that the supercharges are charged under the $R$-current
while they are uncharged under the non-$R$ globally conserved  currents.
The coefficients of the mixing can be read from the variation of the gravitino. In general
they are proportional to the symplectic sections $\mathcal{V}$
defined in (\ref{W-def}). In formulas by referring to  a symplectic vector of charges as $s$ we have $s = t \mathcal{V}$.

The coefficient of the normalization
\footnote{One can choose 
also other normalizations, it is just important to consider this difference when mapping the normalization
of the charges described in supergravity with the ones on the field theory side.}
 is imposed by normalizing the charge of the gravitino to $1$. This choice corresponds to $t =1/ W$ as 
 can be read from the gravitino variation. 
Our $R$ charges become
\begin{equation}
\label{Rchargess}
s = \frac{\mathcal{V}}{W}
\end{equation}
The flavor currents are obtained by acting with derivatives on $\mathcal{V}$.
The combinations of the charges that determines the exact $R$-current  is determined by (\ref{vacuum}). 

At this point we can try to identify the  $\tau_{RR}$-function in AdS$_4$ gauged supergravity.
The coefficient of the flavor two point function in the AdS/CFT correspondence is 
dual to the inverse square Yang Mills coupling in the holographic correspondence.
This observation is the starting point to identify the $\tau_{RR}$-function.
Here we  adapt the discussion of \cite{Barnes:2005bw} to the symplectic invariant formalism.
In this case we can use the matrix
 $\mathcal{M}$
in formula (\ref{Male}) and our starting point becomes the formula
\begin{equation}
\label{conjectural}
\tau_{RR} = \mathcal{T}  \, s^T\mathcal{M}  \bar s
\end{equation}
We observe that this is a real function, depending on a combination of complex $R$-charges.
This notion   requires some interpretations. In a magnetic gauging we can
restrict to the simpler formula $\tau_{RR} \propto \mathcal{I}_{\Lambda \Sigma}  s^\Lambda {\bar s}^\Sigma$.
In this case one can set the imaginary parts of the sections to zero and treat the R-charges as
real. An analogous discussion holds for an electric  gauging.
The situation is more subtle in 
a dyonic gauging. In such a case one should apply a symplectic rotation to identify the correct real combinations leading to the R-charges. 
As discussed in section \ref{section:tau} some dyonic gaugings are more subtle \cite{Guarino:2015jca,Guarino:2015qaa} and deserve further investigations.

We insert the R-charges (\ref{Rchargess})  in  (\ref{conjectural})
and simplify the expression by using the constraints of the special geometry.
In this way we obtain
\begin{equation}
\label{maintau}
\tau_{RR} = \mathcal{T}  \, \mathcal{M} s \bar s 
=\frac{\mathcal{T}  }{|W|^2}
\end{equation}
The extremization condition  corresponds to the requirement of an $\mathcal{N}=2$ 
AdS vacuum (\ref{vacuum}). 
The signs of the second derivatives are imposed from the constraints of the special geometry.
The sign is determined by the  positivity of the scalar metric in (\ref{metric}).
Here we obtain 
\begin{equation}
\partial_{i} \partial_{\bar j} |W|^{-2} \propto  - g_{i \bar j} |W|^{-2},
\end{equation}
This relation does not seem to be correct, because the second derivative of $\tau_{RR}$
should have the opposite sign.
This leads to a maximization of $\tau_{RR}$ and not to a minimization as expected.

Here we discuss a possible resolution of this problem, similarly to the discussion of 
\cite{Barnes:2005bw}, where $\tau_{RR}$ was computed from
the AdS/CFT correspondence.
Here we observe that the function $\tau_{RR}$ at the fixed point
is reproduced by the relation
\begin{equation}
\label{generic}
\tau_{RR} = \mathcal{T} \frac{|W|^{p-2}}{\mathcal{I}_4^{p/4}}
\end{equation}
This extremal value of this function and its first derivative are independent from the value of $p$. 
Notice that the Hessian at the extremum $\partial_i|W|=0$ can be obtained by using  \eqref{HessianW} and it is given by
\begin{eqnarray}
\partial_{\bar j}\partial_i\tau_{RR}&=&g_{i\bar j}(p-2)\tau_{RR}\ .
\end{eqnarray}
Different values of $p$ are consistent with the extremization of the conjectured $\tau_{RR}$-function
but they can lead to a maximization instead of a minimization problems.
We can try to speculate on the origin of the different values of $p$.
In the holographic dictionary the coefficient $\tau_{RR}$, in generic dimensions $d$, is a function of the 
AdS length scale,  $\ell_{AdS}^{d-3}$.
For the case $d=3$ the dependence of $\tau_{RR}$ from $\ell_{AdS}$ drops out.
When we study the behavior of the function $\tau_{RR}$ we vary the scalars in the 
vector multiplets, while keeping the moment maps constant, i.e. the hyperscalars
fixed at their supersymmetric vacuum.
In view of this observation the relation (\ref{moments}) inserts back  
the  AdS scale in the problem, introducing another dynamical object in the minimization problem.
It does not modify the dictionary explained above, but it can modify the derivatives and the 
off-shell behavior of $\tau_{RR}$. 

Another source of mismatch resides in the identification of the $R$-charges.
If we perform a symplectic rotation and reduce to an electric gauging
we can identify the graviphoton with the formula $e^{\mathcal{K}/2} X^\Lambda \mathcal{I}_{\Lambda \Sigma} A^{\Sigma}_\mu$.
The $R$ charges in this case are given by the relation $s^\Lambda = X^\Lambda/(X^\Lambda P_\Lambda)$.
The structure  of the graviphoton is read from the relation above by evaluating 
$\mathcal{I}_{\Lambda \Sigma}$ at the fixed point.
At the fixed point this relation gives the correct mixing of the charges. Out of the fixed point
nevertheless the matrix $\mathcal{I}_{\Lambda \Sigma}$ is a function of the sections. This can be another
source of problems in the identification of $\tau_{RR}$ in (\ref{maintau}).
In the rest of the analysis we will fix $\mathcal{I}_{\Lambda \Sigma}$ to it constant value 
when identifying the charges of the photons.
It would be interesting to come back to this problem.

In the rest of the analysis we fix the coefficient of $p$ above to match the off 
shell relation that one can guess from the AdS/CFT analysis of \cite{Barnes:2005bw}.
Here we propose that the correct power is obtained for $p=6$.
This is a conjectural choice and we will be more concrete on this relation in section \ref{volufree}.
The $\tau_{RR}$-function becomes 
\begin{equation}\label{tau-funct}
\tau_{RR} = \mathcal{T} \frac{|W|^{4}}{\mathcal{I}_4^{3/2}}
\end{equation}
Observe that this discussion is valid in absence of trivial magnetic fluxes.
We will comment on their role in section \ref{zaffa}. Moreover, eq. \eqref{tau-funct}  suggests 
that the functional dependence of $\tau_{RR}$ on the charges is $\tau_{RR}\propto (s^T\mathcal{M} \bar s)^{-2}$.

In this way we have obtained the supergravity dual of the $\tau_{RR}$-minimization principle
discussed in  \cite{Barnes:2005bw}.
We can also fix the proportionality constant $\mathcal{T}$
from the relation (\ref{large}) and from the relation $|W^{min}|^4 = \mathcal{I}_4$. 
We obtain
\begin{equation}
\label{simple}
\tau_{RR}^{min} =  \mathcal{T} \frac{|W^{min}|^{4}}{\mathcal{I}_4^{3/2}}
 = \frac{4}{ \pi^2} F_{S^3}^{max}\quad
  \rightarrow \quad
  \mathcal{T}  = \frac{4}{ \pi^2} F_{S^3}^{max} \sqrt{\mathcal{I}_4}
\end{equation}

By imposing this normalization our candidate  supergravity dual  $\tau_{RR}$-function.
can be written as
\begin{equation}
\label{final}
\tau_{RR} =  \frac{4 F_{S^3}^{max} }{ \pi^2} \frac{|W|^4}{\mathcal{I}_4} 
\end{equation}

We conclude this section by including in the discussion the effect of broken global symmetries.
In 3D there are not constraints coming from the global anomalies but there are
superpotential couplings, that break some of the global symmetries that can mix with the R-charge.
One may wonder if the effect of these couplings can be captured by the $\tau_{RR} $ function.
This idea is borrowed from the 4D case.

In 4D the effects of the broken symmetries
are captured in $a$-maximization by including some Lagrange multipliers in the problem.
The multipliers impose the superpotential (and the anomaly) constraints and they are associated to the coupling constants.
It has been shown \cite{Kutasov:2004xu,Barnes:2004jj} that this procedure matches the perturbative results in field theory, i.e.
one can expand the exact R-charges in terms of the multipliers and match with the perturbative expansion.
In 3D a similar proposal for $\tau_{RR}$ is missing, but it has been shown that the multipliers
can be considered in the extremization problem of $F_{S^3}$ \cite{Amariti:2011xp},
where a two loop matching was observed.
This allows us to think that a similar mechanism can be proposed for $\tau_{RR}$ and 
motivates the search of the supergravity dual of the Lagrange multipliers.

First we have to translate the effect of the broken global symmetries. They are related to the 
presence of massive gauge bosons on the holographic dual side. This effect can be captured in presence of  
hypermultiples.  
So far we only considered the hypermultiplets fixed at their supersymmetric vacuum 
and  we  used them only to 
determine the correct moment maps necessary to identify the mixing of the currents.
Here we turn on some of these fields, considering their role  in the extremization of $\tau_{RR}$.

The effects of the hypermultiplets is obtained by a closer look at the supersymmetric variations.
In the discussion above we have considered only the solution $k^u=0$ to (\ref{vacuum}).
Now we expand around these solutions, and consider some $k^u \neq 0$, at linear order in the hypermultiplets.
These solutions split  $\mathcal{M}_H$ into two parts, $\mathcal{M}_{H_1}$ and $\mathcal{M}_{H_2}$.
For the $\mathcal{M}_{H_1}$ submanifold the situation corresponds to the one described above:
$k^u$ are vanishing and they do not provide further constraints on $\mathcal{V}$. 
Here we concentrate on the submanifold $\mathcal{M}_{H_2}$.
In this case the $n_{H_2}$ non vanishing $k^u$ signal  the spontaneous breaking of the gauge group.
There are $n_{H_2} \leq n_V$ massive gauge bosons, and they acquire the mass by an Higgs mechanism, eating 
$n_{H_2}$ scalars in $\mathcal{M}_{H_2}$, leaving only 
$3 n_{H_2}$ scalars on $\mathcal{M}_{H_2}$.
The moment maps $\mathcal{P}^x$ become functions of the uneaten hyperscalars. Here, as discussed above,
we can use still an $SU(2)$ transformation when expanding around the vacuum, and consider  only one non vanishing component, $\mathcal{P}^3$,
depending on $n_{H_2}$ coordinates on  $\mathcal{M}_{H_2}$.
This is valid if the Killing potentials are expanded around the supersymmetric
vacuum at linear order in the hyperscalars \cite{Tachikawa:2005tq}.
At the same time the solution of the hyperino 
variation for the non vanishing components of $k^u$ impose 
$n_{H_2}$ conditions on the scalars on   $\mathcal{M}_V$.
They are exactly the $n_{H_2}$ constraints imposed by the $n_{H_2}$
hyperscalars in $\mathcal{P}^3$.

This discussion coincides with the one of AdS$_5$ gauged supergravity done in \cite{Tachikawa:2005tq}.
This led to identify the multipliers of the field theory description with the 
hyperscalars in the moment map $\mathcal{P}^3$ enforcing the constraints on 
$\mathcal{V}$.
Here we see the same mechanism at work on the gravity side. 
The presence of the multipliers allows the study of R-symmetric RG flows,
along the lines of \cite{Bertolini:2013vka} (see also \cite{Szepietowski:2012tb} for another discussion on the role of the multipliers
in the AdS$_5$ case).

\section{The $\tau_{RR}$ function of the ABJM model}
\label{section:examples}
In this section we apply our formalism to the calculation of the holographic 
$\tau_{RR}$-function for the ABJM model.

In general the simplest example that one can consider  consists of $\mathcal{N}=2$ gauged supergravity with 
a graviton multiplet and $n_V$ vector multiplets. Here we choose the case 
with $n_V=3$ and focus on a solution discussed in \cite{Cacciatori:2009iz}.
This theory corresponds to a consistent truncation of $S^7$, and it accounts for a
deformation of the ABJM model, along the Cartan $U(1)^4$ of the $SO(8)_R$ symmetry group. 
The model admits a formulation in terms of a prepotential
 \begin{equation}
 \mathcal{F} =- 2 \sqrt{-X^0 X^1 X^2 X^3}
 \end{equation}
The symplectic vector $\mathcal{V}$ and the K\"ahler potential $\mathcal{K}$ are
 \begin{equation}
 \mathcal{V} = e^{\mathcal{K}/2} (1,t_2 t_3,t_1 t_3,t_1 t_2,
 -i t_1 t_2 t_3 ,  -i t_1,    -i t_2 ,
    -i t_3 ) ,
\quad
e^{-\mathcal{K}} =8  Re(t_1) Re(t_2) Re(t_3)
\end{equation}
We consider a purely electric gauging, with charges $g_i = \frac{1}{2}$ ($i=1,\dots,4$). 
The moment maps are constant functions of these charges.
 They can be formulated in terms of $\mathcal{P}^3$ as
\begin{equation}
\mathcal{P}^3 = \frac{1}{2}(0,0,0,0;1,1,1,1)
\end{equation}
The $\tau_{RR}$-function in this case is
\begin{equation}
\tau_{RR} = \frac{ F_{S^3}^{max} }{64\pi^2}
\frac{ |1 +t_1 t_2 +  t_1 t_3 + t_2 t_3|^4}
{  \, (\text{Re}(t_1) \,\text{Re}(t_2) \,\text{Re}(t_3))^2}
 \end{equation}
Where $F_{S^3}^{max}$ corresponds to the maximal value of the $F_{S^3}$, used here as an input, 
obtained from localization.
At the extremal point the scalars $t_i$ are fixed to the $\mathcal{N}=2$
supersymmetric vacuum, $t_i=1$.
This is the supersymmetric attractor mechanism and it corresponds
to the condition (\ref{vacuum}). In the field theory interpretation this corresponds to the extremization 
condition on the R-charges (\ref{Rchargess}).
The ${\tau}_{RR}$ function evaluated at
the minimum  corresponds to the expected result (\ref{large}).

We can also compute the R-charges.
In this case we are in presence of a magnetic gauging and the graviphoton can be written in terms of the sections as
$e^{\mathcal{K}/2} X^\Lambda F^{\mu \nu}_{\Lambda}$.
As discussed above here we describe the charges for the field strength $\mathcal{I}_{\Lambda\Sigma } F^{\Sigma \mu \nu }\equiv F_\Lambda^{\mu\nu}$.
In this sense we treat $\mathcal{I}_{\Lambda \Sigma }$ as a constant matrix.
Moreover the sections can be treated as real. 
We choose the R-charges in this case as $s^\Lambda = e^{\mathcal{K}/{2}} X^\Lambda/2\mathcal{W}$.
We can also turn off the imaginary parts  the coordinates $t_i = b_i + i v_i$ and 
parameterize the R-charges as 
\begin{equation}
\label{chargesABJM}
s^1 = \frac{2}{B}\quad\quad
s^2 = \frac{ 2b_2 b_3 }{B}\quad\quad
s^3 =  \frac{ 2b_1 b_3 }{B}\quad\quad
s^4 =  \frac{ 2b_1 b_2 }{B}
\end{equation}
where we defined $B = 1+\sum_{i<j} b_i b_j$.
At the fixed point the scalars are $b_i =1$ and the R-charges $s^\Lambda$ are all equal to $\frac{1}{2}$.
At $k=1$ the normalization $\tau_{RR}^{min}$ can be extracted from \cite{Jafferis:2011zi}. It is 
\begin{equation}
\tau_{RR}^{min} = 
 \frac{4 F_{S^3}^{max}}{ \pi^2}  =  \frac{4 \sqrt{2} }{3 \pi} N^{3/2} 
\end{equation}
Observe that in this case we can describe the function $\tau_{RR}$ before the extremization as a function of the R-charges
$s^\Lambda$.
We have
\begin{equation}
\label{tauabjm}
\tau_{RR} = 
 \frac{ \sqrt{2} }{3 \pi} N^{3/2} \frac{1}{16 \, s^1 s^2 s^3 s^4}
\end{equation}
It is interesting to compare this result with the expectations from localization.
Indeed in this case we can associate the R-charges to the one of the ABJM model.
This model can be associated to 
a consistent truncation of a deformed ABJM model.
On the field theory side the ABJM model corresponds to a 3D quiver gauge theory with 
$U(N)_{k} \times U(N)_{-k}$ gauge groups, where $k$ is an integer Chern-Simons (CS) level.
There are two pairs of bifundamental fields $a_i$ and $b_j$ with superpotential 
\begin{equation}
\label{WABJM}
W = a_1 b_1 a_2 b_2 - a_1 b_2 a_2 b_1
\end{equation}
The theory has $\mathcal{N}=6$ supersymmetry enhanced to $\mathcal{N}=8$
for $k=1,2$. 
The model that we considered in this section, corresponding to set the FI parameters of the gauging $g_i=1/2$,
is a consistent truncation of the ABJM model preserving the full Cartan $U(1)^4$ symmetry. 
The situation with generic $g_i$ corresponds to the topological twist discussed in 
\cite{Benini:2015eyy}. We will come back to this case in section \ref{zaffa}.

The free energy can be parameterized on the field theory side with the R-charges of the four fields
$a_1$, $a_2$, $b_1$ and $b_2$. The general parameterization respecting 
the  $U(1)^4$ Cartan symmetry is 
\begin{eqnarray}
\Delta_{a_1} = \delta_0 + \delta_1 +\delta_2 + \delta_3, \quad \quad
\Delta_{a_2} = \delta_0 + \delta_1 -\delta_2 - \delta_3\nonumber \\
\Delta_{b_1} =  \delta_0 - \delta_1 +\delta_2 - \delta_3, \quad \quad
\Delta_{b_2} =  \delta_0 - \delta_1 -\delta_2 + \delta_3
\end{eqnarray}
The free energy at large $N$ is 
\begin{equation}
F_{S^3} =  \frac{\sqrt{2} \pi}{3} N^{3/2} \sqrt{\Delta_1 \Delta_2 \Delta_3 \Delta_4}
\end{equation}
The maximization of the free energy fixes $\Delta_{a_i} = \Delta_{b_i}=\frac{1}{2}$.
The exact R-current is obtained by the combination
\footnote{Here we refer to the canonical normalization of the R symmetry $R_0 = (d-2)/2 J_0$, this explains the difference in the normalization  
discussed in (\ref{combo}). }
\begin{equation}
R_{ex} = \delta_0 J_0 + \delta_{i} F_{i} 
\end{equation}
where $F_i$ represent the three $U(1)$ currents that can mix with the R-charge.
In this case, at the fixed point we have $\delta_i$=0.
The canonically normalized  exact R-current  corresponds to $J_0/2$.
This is an expected result, because in this case two of the global $U(1)$s are actually
in the Cartan of the global $SU(2)^2$ symmetry group and  the other $U(1)$ is a baryonic symmetry.

We want to compare this result in terms of the supergravity result obtained in (\ref{tauabjm})
for $\tau_{RR}$.
The graviphoton in this case corresponds to the combination
$s^\Lambda A_{\Lambda}^{\mu}$, where the $s^\Lambda$ are the charges  
parameterized in (\ref{chargesABJM}).
The relation between the $s^\Lambda$ and the $\delta_i$ variables is
\begin{eqnarray}
\label{charges}
\delta_{0} &=&  \frac{s^0+s^1+s^2+s^3}{2},\quad \quad
\delta_{1} = \frac{ s^0+s^1-s^2-s^3}{2} \nonumber \\
\delta_{2} &=& \frac{s^0-s^1+s^2-s^3}{2},\quad \quad
\delta_{3} = \frac{s^0-s^1-s^2+s^3}{2}
\end{eqnarray}
supplemented by the constraint $s^\Lambda P_\Lambda=1$, that here
becomes $\sum s^\Lambda = 2$.
At the vacuum this reproduces the expected relation $\delta_0=1$ and $\delta_i=0$.

To conclude this section we want to compare the functional behavior of the function $\tau_{RR}$ and
the function $F_{S^3}^2 $, in terms 
of the parameterization found in (\ref{charges}).
We obtain the relation
$F_{S^3}^2 (s^\Lambda) \propto \tau_{RR}^{-1}(s^\Lambda)$.
This relation is the one expected from the AdS/CFT correspondence, as we will comment in section \ref{volufree}.
This is consistency check  of our conjecture on the structure  of $\tau_{RR}$.

\section{Relation with the large $N$ $F_{S^3}$  and $Vol(SE_7)$}
\label{volufree}

In this section we comment on the relation between our results and the predictions from localization.
Here we conjectured the structure of $\tau_{RR}$ compatible with  
a functional relation between  $\tau_{RR}$ and $F_{S^3}$ in terms of the
R-charges.
Such a  relation is indeed expected from the results of \cite{Barnes:2005bw}, where the 
relation between $\tau_{RR}$ and the volume form $Vol(Y)$, 
appearing when studying M-theory compactified on a SE$_7$ manifold $Y$, was discussed.

In general $F_{S^3}$ and $\tau_{RR}$ 
are different functions in terms of their
dependence on a generic assignment of the R-charges.
Nevertheless, as we observed in the case of the ABJM theory, our conjectured definition
of  $\tau_{RR}$ leads to the functional relation  $\tau_{RR} \propto F_{S^3}^{-2}$, once the R-charges 
on the two sides of the duality are identified.
This corresponds to the choice $p=6$ in (\ref{generic}).

Let us briefly review the results  of \cite{Barnes:2005bw}.
In AdS/CFT the volume $Vol(Y)$ is parameterized in terms of the Reeb vector $\mathbf{b}$.
This vector is a Killing vector corresponding to one of the $U(1)$ isometries of $Y$. 
This isometry corresponds to the R-symmetry in the dual field theory.
The exact R-charge is obtained by minimizing the volumes
in terms of  the components of $\mathbf{b}$  \cite{Martelli:2006yb}.
It was observed that $\tau_{RR}$, obtained from the KK reduction on the volume formula,
is proportional, at the fixed point, to the inverse volume. This led to an apparent contradiction, 
being both the functions\footnote{Here we refer to the $\tau_{RR}$-function computed from the AdS/CFT correspondence. For this reason we express the dependence from the components of the Reeb vector
$\mathbf{b}$.}  $\tau_{RR}(\mathbf{b})$ 
and $Vol_{\mathbf{b}}(Y)$ minimized by the exact R-charge. The way out discussed in the paper 
was to distinguish a functional dependence of $\tau_{RR}(\mathbf{b})$ from $Vol_{\mathbf{b}}(Y)$
and a normalization to respect of the volume at the fixed point, $Vol_{min}(Y)$. 
In this way it does still make sense to have two different principles of minimization for 
 $\tau_{RR}(\mathbf{b})$ 
and $Vol_{\mathbf{b}}(Y)$ but an inverse functional dependence from the R-charge parametrized by the components of the vector $\mathbf{b}$.
In this paper we observed a similar mechanism at work in gauged supergravity. 
The relation between $\tau_{RR}$  and the volume is \cite{Barnes:2005bw}
\begin{equation}
\label{voltaub}
\tau_{RR}(\mathbf{b}) = \frac{4 \pi^2}{3 \sqrt{6}} \left(\frac{N }{Vol_{min}(Y)}\right)^{3/2}
Vol_{\mathbf{b}}(Y_7)
\end{equation}
On the other hand the general relation between the free energy and the volume $Vol(Y)$ is
\begin{equation}
\label{freeReeb}
F_{S^3}(\mathbf{b}) = N^{3/2} \sqrt \frac{2 \pi^6}{27 Vol_{\mathbf{b}}(Y)}
\end{equation}
By combining the two relations  (\ref{freeReeb}) and (\ref{voltaub})
one obtains
\begin{equation}
\label{general}
\tau_{RR}=
\frac{4}{\pi^2}
\frac{(F_{S^3}^{max})^3}{F_{S^3}^2}
\end{equation}
where $F_{S^3}^{max}$ is the maximized free energy corresponding to $Vol_{min}(Y)$ in 
(\ref{voltaub}).
This leads to a prediction for the relation between $\tau_{RR}$
and the large $N$ free energy $F_{S^3}$.

In other words here we fixed  $p=6$ in (\ref{generic}) to match the predictions
on $\tau_{RR}$ for the ABJM model.
As we discussed above it would be important to have  a more
direct derivation of this result from gauge supergravity.
This may also shed some light on 
a general expectation, based on the analogy with the 4D case \cite{Butti:2005vn}. 
The expectation is that the volume form can be always associated to a quartic function
of the R-charges \cite{Amariti:2011uw,Amariti:2012tj}. 
Gauged supergravity has already been proposed in \cite{Lee:2014rca} to find a similar 
relation. We hope to come back to these problems in the future.

\section{Topological twist and relation with AdS$_2$ BH entropy}
\label{zaffa}
In this section we speculate on some possible relations with a
very interesting result, recently appeared in \cite{Benini:2015eyy}.
In this paper the authors counted the microstates of
asymptotically AdS black holes,
reproducing the Bekenstein-Hawking (BH) entropy from the calculation of an index in the holographic dual gauge theory.
The index is a function of the R-charges of the dual superconformal field theory, and it has been shown that the correct entropy is found once the exact R-charge  is imposed on the index. 
The exact R-charge is obtained by an extremization principle, dual, on the gravity side, 
to an attractor mechanism.
Because of the odd dimensionality the authors proposed a derivation of this extremization in analogy
with the maximization of $F_{S^3}$, observing that the Witten index on $S^1$
has indeed the desired properties. 
Having  $\tau_{RR}$ the same extremization properties of $F_{S^3}$  
one may hope to derive an extremization principle also from our results.
Here we observe that there is a possibility of deriving 
such a relation by reducing to AdS$_2 \times S^2$ the holographic $\tau_{RR}$-function.
When compactifying AdS$_4$ on AdS$_2 \times S^2$ with magnetic fluxes turned on, 
the AdS$_4$ superpotential $W_4$ reduces to the ratio of the AdS$_2$ central charge $Z_{2D}$
and the AdS$_2$  superpotential $W_{2D}$.

As we observed above, a generic function, proportional to the superpotential $|W|$,
is extremized by the exact R-charge. Nevertheless when we consider the presence of magnetic fluxes
they can mix with the R-charge and the various function have different extremization properties.
What seem reasonable is to study the function at $p=2$ in (\ref{generic}). This function is maximized by the exact
R-charge also in the case when the theory is deformed by the fluxes.
Moreover, along the gravitational flow, this function becomes \cite{Dall'Agata:2010gj}
\begin{equation}
\label{relALE}
\tau_{RR} = \frac{2}{\pi G_4 |W_4|^2}
\rightarrow
\frac{1}{G_4} \Big{|} \frac{Z_2}{W_2} \Big{|} 
\end{equation}
During this reduction the scalars cannot be kept fixed and
their mixing provides a different attractor.
In the field theory language the addition of the fluxes is equivalent to 
a topological twist on the flavor symmetries, and the new attractor can 
be reformulated by a different mixing of the R-current
with the fluxes in the dual 1D superconformal quantum mechanics.
 The AdS$_2$ attractor equation in this case fixes the correct mixing and should correspond
to an R-charge extremization principle on the field theory side.
At the vacuum the relation (\ref{relALE}) becomes 
\begin{equation}
\frac{1}{G_4} \Big{|} \frac{Z_2}{W_2} \Big{|}  \propto \frac{R_{S^2}^2}{G_4} \propto S_{BH}
\end{equation}
reproducing the BH entropy.
It would be interesting to investigate in this direction.
For example one can 
try to reproduce the results of 
\cite{Benini:2015eyy}
and further
study the case of other BPS black holes, as the ones
obtained in \cite{Halmagyi:2013sla}
from consistent truncations to AdS$_4$ of 10D $M$-theory.
Our analysis suggests that 
an R-charge extremization principle at work in the dimensional 
reduction of the associated  topologically twisted field theories 
may be captured by the reduction of (\ref{relALE}).

\section{Conclusive discussion: open problems and further investigations} 
\label{section:discussion}

In this paper we studied the coefficient of the two
point function for the R-current of 3D
$\mathcal{N}=2$ SCFTs from $\mathcal{N}=2$ AdS$_4$ gauged supergravity.
Taking advantage of the constraints of the special geometry 
we have conjectured the supergravity dual $\tau_{RR}$-function
(\ref{final}).
We have derived the extremization  principle of  \cite{Barnes:2005bw},
to obtain the exact mixing of the R-current with
the abelian symmetries, in the gravitational setup.
It corresponds to an attractor mechanism for the scalars in the 
vector multiplets.
We discussed also the role of the quaternionic manifold, showing that
the hypermultiplets can  be interpreted 
as Lagrange multipliers constraining the extremization.
The analysis does not
require the existence of a prepotential and it applies for different choices of
gauging in many setups.

In the derivation we conjectured the behavior of $\tau_{RR}$ in order to reproduce the
AdS/CFT predictions in the case of the ABJM model. In this way we obtained 
the relation between the  $\tau_{RR}$-function and the free energy $F_{S^3}$ 
for general R-charges, matching the expectations 
from the volume computations.
More general checks and studies in this direction are necessary.
Here we did not consider other truncations that can have interesting consequences in the 
AdS/CFT correspondence.  For example one can consider 
the  dual theories conjectured in \cite{Hanany:2008fj,Franco:2009sp,Benini:2009qs} for the truncation of $M^{111}$ and  $Q^{111}$  
and compare with the predictions from the volume formula obtained from the geometry.
In this case it would be possible to identify the general behavior of $\tau_{RR}$
in gauged supergravity in terms of the Reeb vector 
for a general truncation of SE$_7$ manifolds along the lines of \cite{Cassani:2012pj}.
Observe that for some of these theories the calculation of the free energy does not 
reproduce the $N^{3/2}$ scaling behavior \cite{Jafferis:2011zi}. Our analysis in gauged supergravity 
may produce a different holographic check for these models.

We also discussed a possible relation between our construction and the 
results of \cite{Benini:2015eyy}. Motivated by this relation we think that it would be interesting 
to perform a direct study of  the R-charge extremization 
problem from the 1D perspective.
This analysis is similar to the ones performed in  \cite{Benini:2012cz,Benini:2013cda,Karndumri:2013iqa}
when flowing to AdS$_5$ to AdS$_3$.

In our analysis we have been interested in cases without 
higher derivatives terms. The inclusion of these contribution should corresponds
to models with non vanishing  TrR. 
Another analysis deserving further investigation is the study of the global properties
of the hyperscalar manifold. This is necessary for studying flows between supersymmetric 
solutions. Here, by including the effects of the Lagrange multipliers,
we restricted to the possibility of R-symmetric supersymmetric RG flows.

Here conclude with a last observation.
The prepotential $\mathcal{F} = C_{IJK} X^I X^J X^K/X^0$
corresponds to the very special K\"ahler geometry and 
it is related to the AdS$_5$ case.
Indeed  it  can be obtained by reducing the  $\mathcal{N} = 2$
AdS$_5$ supergravity.
It would be interesting to understand if this observation has some possible consequences in the relation between the free energy and the central charge obtained in \cite{Giombi:2014xxa}, where an
interpolation between 4D $a$-maximization and 3D $F$-maximization was obtained. 
Recently another connection between 
 $a$ and $F$ was discussed in \cite{Fluder:2015eoa}.
 In this case our analysis may get modified by the presence of
 a dyonic gauging.
One may try to connect these result and
the special geometry of the AdS$_5$ and the AdS$_4$  supergravity.

\section*{Acknowledgments}
It is a pleasure to thank Prarit Agarwal for collaboration during many stages of this project.
We are extremely grateful to Kenneth Intriligator and Alberto Zaffaroni 
for their comments on the draft.
The work of A.G. is supported by the Interuniversity Attraction Poles Programme initiated by the Belgian Science Policy (P7/37), and by COST Action MP1210 "The String Theory Universe".

 \begin{appendix}

\section{ Definitions and useful identities of gauged $\mathcal{N}=2$ supergravity}
\label{appeale}
A general $N=2$ theory\footnote{Definitions and conventions used in the paper are explained in this appendix. For a complete discussion of $N=2$ gauged supergravity we refer to the review\cite{Andrianopoli:1996cm}. For a more general analysis on gauged $N=2$ vacua we refer to \cite{Louis:2012ux} .} can be coupled to $n_V$ vector multiplets $(A_\mu^I,\lambda^{iA},\lambda^{i*}_A,z^i)$, containing complex scalar fields $z^i$ $(I,i=1,..,n_V)$, and $n_H$ hypermultiplets $(\zeta^\alpha,q^u)$, containing real scalars ($\alpha=1,..,2n_H$, $u=1,...,4n_H$).

The $4n_H$ real $q^u$ scalars are in fact coordinates of a quaternionic manifold $\mathcal{QM}$ of quaternionic dimension $n_H$. The choice of gauging considered in this work involves a group of isometries $G\in \mathcal{QM}$. It is defined by a set of moment maps $\mathcal{P}^x(q)$ related to the Killing vectors as \cite{Louis:2012ux}
\begin{eqnarray}
-2k^u_\Lambda K^x_uv&=&\nabla_v\mathcal P^x_\lambda\ ,
\end{eqnarray}
where $K^x$ is the curvature of the $SU(2)$ connection on the quaternionic manifold.

 The complex scalars $z^i$ parametrize a special K\"ahler manifold $\mathcal{SM}$, whose geometry is completely defined by a K\"ahler potential $K(z,\bar z)$, from which the metric of the manifold is derived as
\begin{eqnarray}
\label{metric}
g_{i\bar j}(z,\bar z)&=&\partial_i\partial_{\bar j}K(z,\bar z)\ .
\end{eqnarray}
It is convenient to parametrize the special K\"ahler scalar fields with holomorphic symplectic sections of a projective bundle, $( X^{\Lambda}(z), F_\Lambda(z))^T$ , $\Lambda=0,1,..,n_V$, satisfying (bar indicates complex conjugation)
\begin{eqnarray}
F_\Lambda \bar X^{\Lambda}-X^{\Lambda}\bar F_\Lambda=-ie^{-K}\ .
\end{eqnarray}
The expression above defines the symplectic product as
\begin{eqnarray}
\langle A_1,A_2 \rangle=A_1^T\Omega A_2\ ,\qquad \Omega=\left(
\begin{array}{cc}
0&I_{2n_V+2}\\-I_{2n_V+2}&0
\end{array}
\right)\ ,
\end{eqnarray}
on any $Sp(2n_v+2)$ vector $A=(A_\Lambda,A^\Lambda)$. The normalized symplectic sections are then 
\begin{eqnarray}
\mathcal{V}&=& e^{K/2}(X^{\Lambda}(z), F_\Lambda(z))^T\ , 
\qquad  \langle \mathcal{V}, \bar{\mathcal{V}} \rangle=-i\ ,
\end{eqnarray}
they satisfy 
\begin{eqnarray}
D_i \mathcal{V}&=&\partial_i \mathcal{V}+\frac12\partial_i K\, \mathcal{V}\ ,\qquad 
D_{\bar i}\mathcal{V}=\partial_{\bar i} \mathcal{V}-\frac12\partial_{\bar i} K\, \mathcal{V}=0\ ,
\nonumber\\
D_{\bar i} \bar{\mathcal{V}}&=&\partial_{\bar i} \bar{\mathcal{V}}+\frac12\partial_{\bar i} K\, \bar{\mathcal{V}}\ ,\qquad 
D_{ i}\bar{\mathcal{V}}=\partial_i \bar{\mathcal{V}}-\frac12\partial_i K\, \bar{\mathcal{V}}=0\ .
\end{eqnarray}
By using the special geometry identity
\begin{eqnarray}
D_{\bar j}D_i\mathcal{V}=g_{i\bar j}\mathcal{V}\ ,
\end{eqnarray}
one can derive the following relations used in Sec.\ref{section:main}
\begin{eqnarray}\label{HessianW}
\frac{2 \partial_{\bar j}\partial_i|W|}{|W|}\big|_{\partial_i|W|=0}&=&g_{i\bar j}\big|_{\partial_i|W|=0}\ , \nonumber\\ 
\partial_{\bar j}\partial_i |W|^{p-2}\big|_{\partial_i|W|=0}&=&g_{i\bar j}(p-2)|W|^{p-2}\big|_{\partial_i|W|=0}\ .
\end{eqnarray}

\end{appendix}

\bibliographystyle{JHEP}
\bibliography{BibFile}

\end{document}